%% file: manuscript.tex
\documentclass[conference]{IEEEtran}
\IEEEoverridecommandlockouts
\usepackage{cite}
\usepackage{amsmath,amssymb,amsfonts}
\usepackage{algorithmic}
\usepackage{graphicx}
\usepackage{textcomp}
\usepackage{xcolor}
\usepackage{color,soul}

\input{mydefinitions}

\def\BibTeX{{\rm B\kern-.05em{\sc i\kern-.025em b}\kern-.08em
    T\kern-.1667em\lower.7ex\hbox{E}\kern-.125emX}}
\begin{document}

\title{Highly Directional Scattering of Terahertz Radiation by Cylinders using Complex-Frequency Waves}

\author{
\IEEEauthorblockN{Iridanos Loulas}
\IEEEauthorblockA{\textit{Section of Cond. Matter Physics} \\ \textit{Dept. of Physics} \\
\textit{Nat. Kapod. Univ. of Athens (NKUA)}\\
Athens, Greece \\
iridanos@phys.uoa.gr}
\and
\IEEEauthorblockN{Grigorios P. Zouros}
\IEEEauthorblockA{\textit{Section of Cond. Matter Physics} \\ \textit{Dept. of Physics} \\
\textit{Nat. Kapod. Univ. of Athens (NKUA)}\\
Athens, Greece \\
zouros@phys.uoa.gr, zouros@ieee.org}
\and
\IEEEauthorblockN{Evangelos Almpanis}
\IEEEauthorblockA{\textit{Section of Cond. Matter Physics} \\ \textit{Dept. of Physics} \\
\textit{Nat. Kapod. Univ. of Athens (NKUA)}\\
Athens, Greece} 
\IEEEauthorblockA{\textit{Institute of Nanoscinence and Nanotechnology} \\
\textit{Nat. Cent. Sci. Research DEMOKRITOS}\\
Athens, Greece \\
ealmpanis@phys.uoa.gr, ealmpanis@gmail.com}
\and
\IEEEauthorblockN{~~~~~~~~~~~~~~~~~~~~~~~~~~~~~~~~~~~~~~~~~~~Kosmas L. Tsakmakidis}
\IEEEauthorblockA{~~~~~~~~~~~~~~~~~~~~~~~~~~~~~~~~~~~~~~~~~~~~~~~~\textit{Section of Cond. Matter Physics} \\ ~~~~~~~~~~~~~~~~~~~~~~~~~~~~~~~~~~~~~~~~~~~~~~~~\textit{Dept. of Physics} \\
~~~~~~~~~~~~~~~~~~~~~~~~~~~~~~~~~~~~~~~~~~~~~~~\textit{Nat. Kapod. Univ. of Athens (NKUA)}\\
~~~~~~~~~~~~~~~~~~~~~~~~~~~~~~~~~~~~~~~~~~~~~~~~~Athens, Greece \\
~~~~~~~~~~~~~~~~~~~~~~~~~~~~~~~~~~~~~~~~~~~~~~~~~ktsakmakidis@phys.uoa.gr}
}

\maketitle

\begin{abstract}
In this study we investigate the directional scattering of terahertz radiation by dielectric cylinders, focusing on the enhancement of directionality using incident radiation of complex-frequency. We explore the optimization of the second Kerker condition, which corresponds to backward scattering. At first, by carefully tailoring the electric and magnetic polarizabilities of the cylinders, we successfully achieve significant backward scattering, and then manage to even further improve it by deploying a decaying incoming wave (\textit{complex}-frequency). Additionally, we present preliminary results on the directional scattering of THz radiation by a magneto-optical cylinder, demonstrating the potential of this approach for advanced control over the propagation of THz waves. Our findings contribute to a deeper understanding of THz directional scattering and pave the way for the development of novel THz devices and applications, such as high-resolution imaging, sensing, and communication systems.
\end{abstract}

\begin{IEEEkeywords}
Complex-frequency, Kerker scattering, terahertz radiation.
\end{IEEEkeywords}

\section{Introduction}

Historically, electromagnetic waves in the 0.1–10~THz frequency regime, so-called terahertz band, were hard to generate, manipulate, and detect. These problems were caused by the absence of adequate sources and detectors, the high absorption of THz radiation by most materials, and the complexity of designing and constructing terahertz-operating components. Recent technological developments have advanced THz research, bringing up several possible uses in diverse fields, such as communications, sensing, imaging and spectroscopy. 

In most of the above mentioned areas, it is essential to achieve directionality of the THz radiation. For instance, in wireless communication systems, directional transmission and reception of THz waves can reduce signal interference and improve communication quality. Furthermore, in sensing, directional THz sensors can improve the sensitivity and selectivity of measurements by focusing the incident radiation on a particular area of interest, or by isolating the response from specific materials or objects. This can lead to more accurate and reliable results.
In THz imaging systems, directionality allows for better spatial resolution and depth information by focusing the radiation on a specific area. This can help to minimize scattering, reduce noise, and improve image quality. 
In THz spectroscopy, directionality can be used to increase the signal strength and improve the measurement accuracy by concentrating the radiation on the sample under investigation. This can lead to higher spectral resolution and better detection of specific molecular fingerprints or signatures, allowing for more accurate identification and analysis of materials.

A possible way to  obtain directionality is through the so-called Kerker scattering \cite{gef_et_al_12}. The concept of Kerker scattering involves two conditions, known as the first and second Kerker conditions \cite{gef_et_al_12}, which lead to either forward or backward scattering, respectively. However, in passive dielectric spheres and cylinders, the achieved directionality for a monochromatic beam is not perfect. To overcome this problem, it was recently proposed~\cite{kim2022beyond} that by illuminating a particle with radiation of complex-frequency~\cite{tsakmakidisbook,tsakmakidisprl2014,kirbyprb2011,tsakmakidisscadv2018}, i.e., a sinusoidally decaying wave, it is possible to achieve perfect directional scattering. 

To that end, we study the enhancement of directional scattering of THz radiation by dielectric cylinders using incident radiation of complex-frequency, and optimizing the second Kerker condition. We employ a rigorous analytical frequency-domain approach and validate our results with finite-difference time-domain (FDTD) simulations. Finally, we  explore a potential future direction for advanced control over THz wave propagation in magneto-optical systems that exhibit directional scattering over different angles.

\begin{figure}[!t]
\centerline{\includegraphics[scale=1.6]{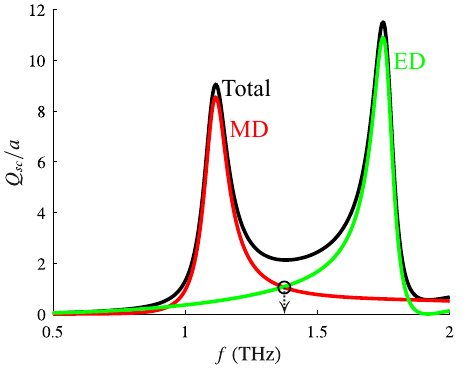}}
\caption{Normalized total scattering cross section spectrum $Q_{sc}/a$ of a dielectric cylinder with radius $a=20~\mu$m. Black curve: total scattering cross section, i.e., including all terms of Eq.~\eqref{4} up to a truncation order that ensures convergence;
red curve: MD contribution to the total scattering cross section; green curve: ED contribution to the total scattering cross section. Black circle: second Kerker point at $f=1.37$~THz.}
\label{geometry}
\end{figure}

\section{Method of calculation}\label{MOC}

At first, we study an infinitely long cylinder of circular cross-section located in free space. This scatterer has radius $a$ and consists of a non-magnetic high-index isotropic dielectric material characterized by a permittivity $\epsilon=\epsilon_r\epsilon_0$ and permeability $\mu_0$, where $\epsilon_0$ and $\mu_0$ are the vacuum values. The cylinder is illuminated by a TE plane-wave impinging normally from the negative values of $x$ towards the positive values of $x$. Adopting the $\exp(i\omega t)$ time dependence, the scattered electric far-field can be expressed as
\balg{1}
{\mathbf E}^{sc}(r,\varphi)\sim E_0\frac{e^{-ik_0 r}}{\sqrt{r}}{\mathbf f}(\varphi),\quad r\rightarrow\infty,
\ealg
where $E_0$ is the amplitude of the incident plane-wave, $k_0$ is the wavenumber of free space, and ${\mathbf f}(\varphi)$ is the scattering amplitude, given by
\balg{2}
\!\!\!\!{\mathbf f}(\varphi)\!\!=\!\!\sqrt{\frac{2i}{\pi k_0}}\!\!\sum_{m=-\infty}^\infty\!\! i^m e^{-im\varphi} \Big(ia_m\hat{\varphi}+b_m\hat{z}\Big)\!\!\equiv\!\!\frac{1}{\sqrt{k_0}}\hat{{\mathbf f}}(\varphi).
\ealg
In \eqref{2}, $a_m$ and $b_m$ are the expansion coefficients of the scattered electric field. For TE scattering, ${\mathbf f}(\varphi)=f_\varphi(\varphi)\hat{\varphi}$ and $\hat{{\mathbf f}}(\varphi)=\hat{f}_\varphi(\varphi)\hat{\varphi}$. Then, the scattering width $\sigma(\varphi)$, $0\leqslant\varphi<2\pi$, can be calculated by its definition via \cite{balanis}
\balg{3}
\sigma(\varphi)=\lim_{r\rightarrow\infty}2\pi r\frac{|{\mathbf E}^{sc}(r,\varphi)|^2}{|{\mathbf E}^{inc}(r,\varphi)|^2}=\frac{2\pi}{k_0}|\hat{f}_\varphi(\varphi)|^2,
\ealg
where ${\mathbf E}^{inc}(r,\varphi)$ is the electric field of the incident plane-wave. Finally, the total scattering cross section, for TE scattering, is given by
\balg{4}
Q_{sc}=\int_0^{2\pi}|f_\varphi(\varphi)|^2{\rm d}\varphi=\frac{4}{k_0}\sum_{m=-\infty}^\infty |a_m|^2.
\ealg

The magnetic dipolar (MD) contribution to the total scattering cross section $Q_{sc}$ is obtained by keeping only the $m=0$ term in \eqref{4}. In addition, the electric dipolar (ED) and the  electric quadrupolar (EQ) contributions to the $Q_{sc}$ are obtained by keeping the $m=\pm1$ and the $m=\pm2$ terms, respectively, in \eqref{4}. Such a multipolar decomposition, which is inherit in this analytical approach, allows for the determination of Kerker points~\cite{gef_et_al_12} for directional scattering, as it is discussed below.

The analytical frequency-domain approach has also the advantage to incorporate directly incident waves of complex-frequencies~\cite{tsakmakidisbook,tsakmakidisprl2014,kirbyprb2011,tsakmakidisscadv2018,kim2022beyond}, placing a complex $\omega$ as an argument. On the other hand, time-domain simulation of complex-frequency waves requires input signals with temporally growing or attenuating behaviors~\cite{kim2022beyond,guan2023overcoming,tsakmakidisbook,tsakmakidisprl2014,kirbyprb2011,tsakmakidisscadv2018}. In order to perform time-domain simulations for comparison reasons, we employ MEEP (MIT Electromagnetic Equation Propagation), a free, open-source software package for simulating electromagnetic systems under the FDTD method~\cite{oskooi2010meep}. 

\begin{figure}[!t]
\centerline{\includegraphics[scale=1.6]{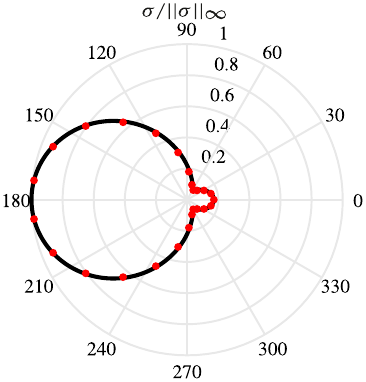}}
\caption{Normalized scattering width $\sigma/||\sigma||_\infty$ versus the observation angle $\varphi\in[0,2\pi)$ at $f=1.37$~THz of the second Kerker point of \fig{geometry}. Black curve: analytical solution; red dots: MEEP.}
\label{fig:real}
\end{figure}

\section{Numerical Results}

We assume an isotropic high-index dielectric cylinder with circular cross section of radius $a=20~\mu$m, located in air, with permittivity $\epsilon=25\epsilon_0$. Figure~\ref{geometry} depicts the normalized total scattering cross section $Q_{sc}/a$ versus the frequency $f$ of the incident plane-wave. The incident wave is assumed TE-polarized. The $Q_{sc}/a$ (black curve) is calculated accounting for all spherical harmonics and angular momentum terms necessary for convergence of Eq.~\eqref{4}. Additionally, the MD (red curve) and the ED (green curve) terms are separately depicted. As clearly indicated in Fig.~\ref{geometry}, the resonant peak appearing close to $1.1$~THz corresponds to the MD resonance, while the higher-frequency one close to $1.75$~THz corresponds to the ED resonance. The frequency pointed out with an arrow, where the ED and MD contributions intersect, corresponds to the second Kerker point which indicates strong back-scattering of the incident radiation~\cite{gef_et_al_12,kim2022beyond}. The frequency of this Kerker point is $f=1.37$~THz.

The normalized scattering width $\sigma/||\sigma||_\infty$ of this second Kerker point, over all observation angles $\varphi$, is shown in Fig.~\ref{fig:real}. Our analytical calculation, shown with black curve, is also reproduced by MEEP package, as shown with the red dots on the same graph. As it is obvious from this polar plot, the back-scattering ($\varphi=180^\circ$) is not perfect, since a small residual at the forward direction ($\varphi=0^\circ$) exists. Therefore, the question that arises is whether or not frequencies exist that completely suppress the forward-scattering. 

To answer this, we first observe that the scattering width $\sigma(\varphi)$ in Eq.~\eqref{3} is also a function of frequency $f$ via the wavenumber $k_0=2\pi f\sqrt{\epsilon_0\mu_0}$. We therefore write $\sigma=\sigma(\varphi,f)$, we fix the forward-scattering observation angle to $\varphi=0^\circ$, and seek for solutions $\bar{f}$ such that $\sigma(0^\circ,\bar{f})$=0. Actually, the suppression of the forward-scattering occurs for a complex-frequency. To find it, we search the complex plane for possible complex roots of $\sigma(0^\circ,\bar{f})$, in the vicinity of Kerker frequency $f=1.37$~THz. With the aid of an appropriate complex roots finding algorithm \cite{zou_18}, we find that this suppression occurs for the complex-frequency $\bar{f}=1.254+i0.129$~THz, which means that a plane-wave with this complex-frequency would excite this totally-back-scattering mode.

\begin{figure}[!t]
\centerline{\includegraphics[scale=1.6]{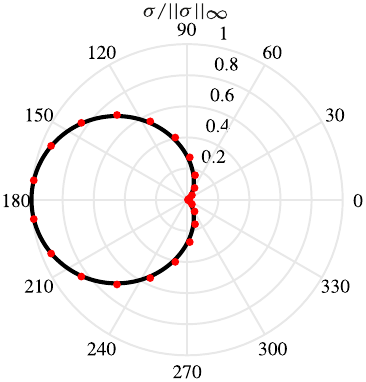}}
\caption{Normalized scattering width $\sigma/||\sigma||_\infty$ versus the observation angle $\varphi\in[0,2\pi)$ at the complex-frequency $\bar{f}=1.254+i0.129$~THz that suppresses the forward scattering width at $\varphi=0^\circ$. Black curve: analytical solution; red dots: MEEP.}
\label{fig:complex}
\end{figure}

In Fig.~\ref{fig:complex} we assume an incident plane-wave with complex-frequency $\bar{f}=1.254+i0.129$. The analytical calculation of the normalized scattering width is shown with a black curve, while the respective FDTD simulation assuming a decaying wave with central frequency ${f}=1.254$~THz and decay rate ${\rm Im}~\omega=0.811\times10^{12}$~rad/s, is depicted with red dots in the same graph. We observe that the forward scattering residual of the monochromatic illumination shown in Fig.~\ref{fig:real}, is now totally suppressed. This particular response can be explained through the concept of ``virtual gain'' which is achieved when specific resonances are activated by exponentially decaying signals at tailored complex-frequencies. The cylinder releases energy stored from earlier cycles where the input signal was stronger due to its decay, and at the specific complex-frequency, the electric and magnetic dipoles reach a quasi-stationary state that cannot be achieved by passive scatterers under monochromatic excitation, but can be achieved here because the dipoles are radiating stored energy. In this state, the dipoles cancel each other out in the forward direction but constructively interfere in the backward direction. We point out, however, that the imaginary part of the complex-frequency must be small enough to avoid any instabilities caused by the virtual gain~\cite{tsakmakidisbook,tsakmakidisprl2014,kirbyprb2011,tsakmakidisscadv2018,kim2022beyond}.

\begin{figure}[!t]
\centerline{\includegraphics[scale=1.6]{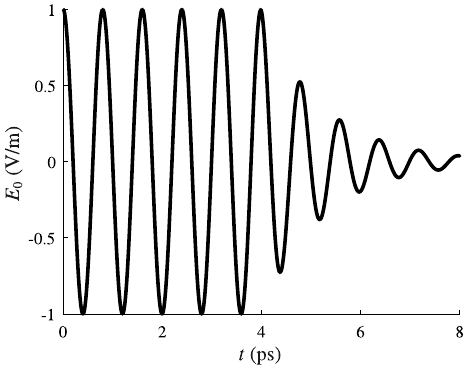}}
\caption{The time-domain excitation pulse of the incident plane-wave in MEEP.}
\label{fig:pulse}
\end{figure}

As already mentioned above, the complex-frequency is inserted directly in the analytical solution, however, in FDTD it is inserted in terms of an exponentially decaying wave, i.e., $E(t) \propto e^{i \bar{\omega} t}$, where $\bar{\omega}=2\pi\bar{f}$ is the complex angular frequency.
%
%
We note that although the idea of an ideal complex-frequency wave is mathematically possible, care needs to be exercised because for time approaches negative infinity, the energy would become infinite. In this respect, to make the complex-frequency wave excitation physically plausible, it is necessary to apply a truncation at a time $\tau_0$, i.e., we start by exciting the scatterer with a
monochromatic signal, then convert to a complex-frequency excitation at $t = \tau_0$. The characteristic input wave, used for the FDTD simulation of Fig.~\ref{fig:complex}, is shown in Fig.~\ref{fig:pulse}. Until a time $\tau_0=4$~ps, the wave is monochromatic, while for $\tau_0>4$~ps, the wave decays exponentially with a decay rate of ${\rm Im}~\omega=0.811\times10^{12}$~rad/s. 


\begin{figure}[!t]
\centerline{\includegraphics[scale=1.6]{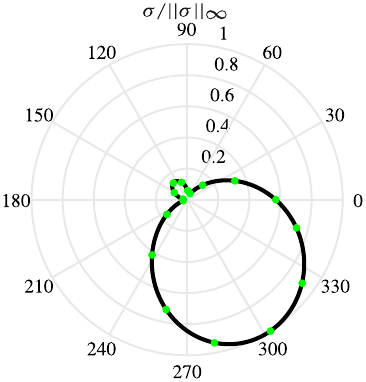}}
\caption{Normalized scattering width $\sigma/||\sigma||_\infty$ versus the observation angle $\varphi\in[0,2\pi)$, for the core-shell cylinder illuminated at $f=2.017$~THz, under $B_0=0.1$~T. The scattering is directed towards $\varphi=300^\circ$, i.e., it ``bends'' $60^\circ$ with respect to the incident direction. Black curve: volume integral equation method; green dots: COMSOL.}
\label{fig:b0_real}
\end{figure}

 Finally, we consider such a dielectric cylinder coated with a gyroelectric magneto-optical medium with amplifying properties. The gain is used instead of a decaying time-domain excitation. In particular, the core-shell structure consists of an isotropic dielectric core of permittivity $\epsilon=25\epsilon_0$ and radius $a_1=14~\mu$m, and an active gyroelectric shell of outer radius $a_2=20~\mu$m magnetized along the $z$-axis, by applying an external magnetic flux density ${\bf{B}}=B_0 {\bf{z}}$. In view of this, the anisotropic permittivity tensor is expressed in Cartesian coordinates by $\e(B_0)=\epsilon_1(B_0)(\ux\ux^T+\uy\uy^T)+i\epsilon_2(B_0)(\ux\uy^T-\uy\ux^T)+\epsilon_3\uz\uz^T$, where $T$ denotes transposition. The precise values of $\epsilon_1(B_0)$, $\epsilon_2(B_0)$ and $\epsilon_3$ are given in \cite{zouros2021three}; these are the same as that of indium antimonide (InSb), a semiconductor with unique magneto-optical properties \mbox{\cite{tsa_she_sch_zhe_uph_den_alt_vak_boy_17}}, where we have reversed the sign of losses to turn it to an amplifying medium. To solve this problem, we employ a volume integral equation method \cite{kat_zou_rou_21}. In \fig{fig:b0_real} we illustrate that, at a specific frequency of $2.017$~THz, the scattering is directed towards $\varphi=300^\circ$, i.e., it ``bends'' $60^\circ$ with respect to the incident direction. This effect is due to higher-order coupling between a resonant MD mode and the almost resonant ED and EQ modes, observed under an external bias of $B_0=0.1$~T and for $a_1/a_2=0.7$. The result from the volume integral equation method (black curve in \fig{fig:b0_real}) is confirmed by a finite-element method simulation (green dots in \fig{fig:b0_real}) using COMSOL. When the external radius $a_2$ is kept fixed and the ratio $a_1/a_2$ changes, ``bends'' with different angles in the scattering width are observed for different values of $B_0$. This allows us to tailor the directional scattering at desirable directions. Although the results in this core-shell configuration are preliminary, our ongoing research focuses on the use of complex-frequencies for eliminating the scattering at the opposite direction than that the radiation ``bends'' (for example, in the plot of \fig{fig:b0_real} at the direction of $120^\circ$), or to direct the scattering at angles that are not possible to achieve using real frequencies.


\vspace*{-2mm}

\section{Conclusions}

In this study, we investigated the directional scattering of terahertz radiation by dielectric cylinders, focusing on enhancing directionality using incident radiation of \textit{complex}-frequency. By employing a rigorous analytical frequency-domain approach, and verifying our results with finite-difference time-domain simulations, we were able to optimize the second Kerker condition corresponding to backward-scattering: We achieved considerable backward-scattering by carefully tailoring the electric and magnetic polarizabilities of the cylinder, and further improved it by assuming a decaying incoming wave (complex-frequency).

Additionally, our preliminary results on the directional scattering of THz radiation by a magneto-optical cylinder demonstrate the potential of this approach for advanced control over the propagation of THz waves. Our findings contribute to a deeper understanding of directional scattering in the THz frequency regime and pave the way for the development of novel THz devices and applications, such as high-resolution imaging, sensing, and communication systems.  

\vspace*{-2mm}

\section*{Acknowledgement}

The authors acknowledge support for this research by the General Secretariat for Research and Technology (GSRT) and the Hellenic Foundation for Research and Innovation (HFRI) under Grant No. 4509.

\vspace*{-2mm}

%
%

\end{document}

%% file: mydefinitions.tex
\newcommand{\fig}[1]{Fig.~\ref{#1}}

\newcommand{\bea}{\begin{eqnarray}}
\newcommand{\beal}[1]{\begin{eqnarray}\label{#1}}
\newcommand{\eea}{\end{eqnarray}}
\def\balg#1#2\ealg{\begin{align}\label{#1}#2\end{align}}
\def\balgnl#1\ealgnl{\begin{align*}#1\end{align*}}

\newcommand{\e}{\boldsymbol{\epsilon}}

\newcommand{\ux}{{\mathbf e}_x}
\newcommand{\uy}{{\mathbf e}_y}
\newcommand{\uz}{{\mathbf e}_z}